\documentclass[fleqn,twoside]{article}
\usepackage{espcrc2}
 
\usepackage{graphicx}
\usepackage[figuresright]{rotating}
 
\newcommand{\AmS}{{\protect\the\textfont2
  A\kern-.1667em\lower.5ex\hbox{M}\kern-.125emS}}
\newcommand{\be}{\begin{eqnarray}}
\newcommand{\ee}{\end{eqnarray}}

\newcommand{\dm}{\mbox{$\Delta m_{21}^2$~}}

\newcommand{\kl}{\mbox{KL~}}

\newcommand{\sss}{\sin^2 \theta_{12}}

\title{Global Analysis of Neutrino Oscillation}
 
\author{Srubabati Goswami\address{ Harish-Chandra Research Institute,
        Chhatnag Road, Jhusi,\\
        Allahabad  211 019, INDIA}
        \thanks{Presented the talk at Neutrino 2004}
        Abhijit Bandyopadhyay\address{Theory Group, Saha Institute of
                                        Nuclear Physics , \\
                   1/AF, Bidhannagar, Calcutta 700 064, INDIA }
        and
        Sandhya Choubey\address{INFN, Sezione di Trieste and
               Scuola Internazionale Superiore di Studi Avanzati,\\
                 I-34014,
                 Trieste, Italy}
                }
\begin{document}

\begin{abstract}
We present the constraints on neutrino oscillation 
parameters $\Delta m^2_{\odot}$ and $\theta_{\odot}$ 
governing the solar neutrino oscillations 
from two generation analysis of 
solar and KamLAND  data.  
We include the latest 766.3 ton year KamLAND data in our analysis. 
We also present the allowed values of parameters 
$\Delta m^2_{atm}$ and $\sin^2\theta_{atm}$ 
from two generation oscillation analysis of SuperKamiokande
atmospheric and K2K data. 
For both cases we discuss the precision achieved in the present 
set of experiments and also how the precision can be improved 
in future. 
We also obtain the bounds on $\theta_{13}$  
from three generation 
analysis of global oscillation data. 
We emphasise on  the roles played by different data sets in  
constraining 
the allowed parameter ranges. 
\end{abstract}

\maketitle

\section{Introduction}
Compelling evidence in favour of neutrino 
oscillation first came from 
atmospheric neutrino flux measurements at SuperKamiokande (SK) \cite{sk98}. 
The   
solar neutrino shortfall at  
SAGE, Gallex,
GNO \cite{sol} ,
Kamiokande and SK \cite{SKsolar} 
were suggestive of neutrino oscillation. 
This was firmly  established by SNO
\cite{sno1,sno2} 
and KamLAND (KL) \cite{kl162,kl766}. The later  
provided the first evidence of oscillation
of reactor neutrinos. 
The K2K experiment 
demonstrated  oscillations of accelerator 
neutrinos \cite{k2k}. Earlier the 
LSND \cite{lsnd} experiment reported positive evidence of 
$\nu_\mu$ ($\bar{\nu_\mu}$) oscillation using low energy 
accelerators though this 
was not confirmed by the similar experiment KARMEN. 
The MINIBOONE experiment will provide an independent check 
\cite{miniboone}. 

Aim of global oscillation analysis is to analyze the experimental data using a 
suitable statistical procedure and extract the information 
on neutrino oscillation parameters 
$\Delta m^2$ (mass squared difference in vacuum) and
$\theta$ (mixing angle in vacuum).
Various statistical methods have been adopted  
starting from the standard covariance method,
pull method \cite{lisipull},Frequentist method \cite{freq} 
and Bayesian Analysis 
\cite{bay}. 

\section{Solar Neutrino Oscillation parameters}
\begin{figure}[htb]
\includegraphics*[width=20pc]{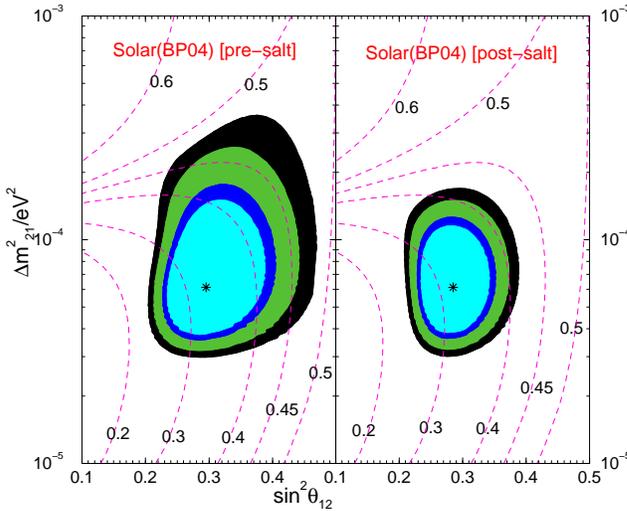}
\caption{The 90\%, 95\%, 99\% and 99.73\% C.L.
allowed regions in the $\dm-\sin^2\theta_{\odot}$ plane from global
$\chi^2$-analysis of solar neutrino  data.
Also shown are the iso-CC/NC contours.
}
\label{sol2osc}
\end{figure}

In this section we discuss the constraints on solar neutrino oscillation 
parameters $\Delta m^2_{\odot} \equiv
\Delta
m^2_{21}$, $\theta_{\odot} \equiv \theta_{12}$ from two flavour
$\nu_e - \nu_{\rm active}$ 
 analysis. 
\\ \\
\underline{{\bf Allowed area 
from global Solar Data  
}}:
In our analysis of  global solar data we 
include the total rates from the radiochemical experiments
Cl and Ga (Gallex, SAGE and GNO combined) \cite{sol} and
the 1496 day SK Zenith
angle spectrum data \cite{SKsolar}, the SNO spectrum data from the 
pure 
$D_2O$ phase \cite{sno1}, as well as 
the CC, NC and ES rates from the salt phase
\cite{sno2}. 
We use {BP04} fluxes \cite{bp04}
 and keep the $^{8}{B}$ flux normalisation free. 
Further details of our 
analysis can be found in  
\cite{snocc,snonc,snosaltus}.
In the left panel of Figure \ref{sol2osc}
we show the allowed region in the parameter space
including the SNO spectrum data
from the pure $D_2O$ phase while the right panel 
shows the allowed regions after including the data from the salt phase. 
The best-fit values of parameters obtained are 
\\
$\bullet$
$\dm$=6.1$\times10^{-5}$eV$^2$,
$\sss$ = 0.3,
$f_B$=0.89.\\ 
$\bullet$99\% C.L.  range {(before salt)} \\
$\dm$=(3.1-25.7) $\times 10^{-5}$eV$^2$,
$\sss$=0.21 -0.44 \\
$\bullet$99\% C.L.  range {(after salt)} \\
$\dm$=(3.2-14.8)$\times 10^{-5}$eV$^2$,
{$\sss$}=0.22 -0.37. \\
The allowed region corresponds to Large-Mixing-Angle (LMA)
MSW solution.
The impact of the salt data is to 
tighten the upper limit on {$\dm$} and {$\sss$}.   
The iso-CC/NC curves in Fig. \ref{sol2osc}
shows that for higher values of $\sss$ and
$\dm$ the 
$CC/NC$ value is higher.  Since the {CC/NC} ratio for the
salt phase data is 0.31 as opposed to 0.35 in the pure $D_2O$ phase 
the allowed regions shift to smaller $\sss$ and $\dm$ values 
following the iso-{CC/NC} contours.
\\ \\
\underline{\bf Allowed area from KamLAND spectra}:
\begin{figure}[htb]
\includegraphics[width=8cm,height=6cm,clip=]{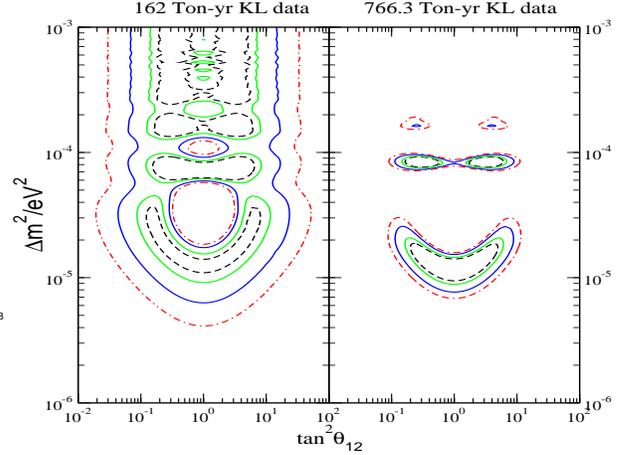} 
\caption{Same as in figure 1 but in 
in the $\dm-\tan^2\theta_{12}$ plane from global
$\chi^2$-analysis of \kl spectral data.
}
\label{klspec}
\end{figure}
We next study the impact of the 
new results from the \kl
experiment in constraining the oscillation parameters. 
%
The survival probability for \kl is :
\be
P
_{\bar{e}\bar{e}}^{KL} =
1 - \sin^22\theta_{12}\sin^2\left(\frac{1.27\Delta m_{21}^2 L}
{E_{\nu}}\right)
\ee
 neglecting the small matter effect for lower values of $\Delta m^2_{21}$. 
Assuming CPT conservation i.e same oscillation parameters for
neutrinos and antineutrinos \kl is sensitive to 
$\dm$ in the  
LMA region. 

\begin{figure}
\includegraphics[width=16pc,clip=]{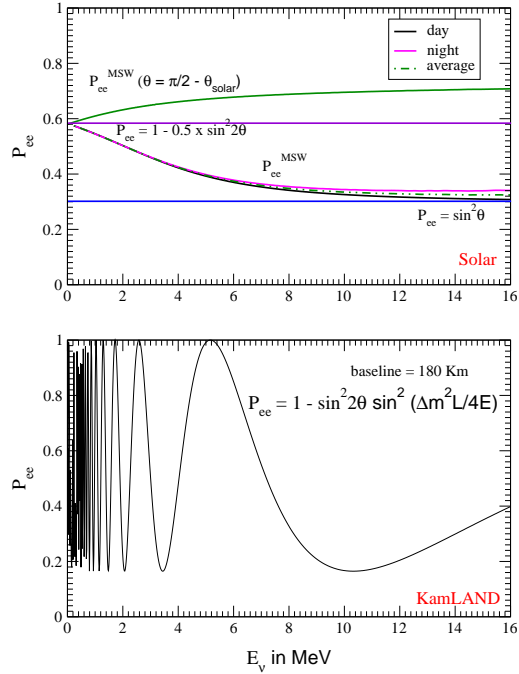}   
\caption{\label{prob} 
The probabilites for solar and \kl as a function of energy}
\end{figure}

In  the right panel of 
fig \ref{klspec} we show the allowed regions obtained 
after including the latest \kl data in our analysis while the left panel 
shows the allowed regions without this data. 
The figure shows that 
higher $\Delta m^2$ regions, which correspond to an averaged survival 
probability and hence flat spectra, 
reduce in size due to the increased distortion in the  
new \kl spectrum data \cite{kl766}. 
For details of our \kl analysis we refer to \cite{kl766us,kl162us}.
\\ \\
\underline{\bf Allowed area from Solar+KamLAND}:  
\begin{figure}[htb]
\includegraphics[width=8cm,height=6cm,clip=]{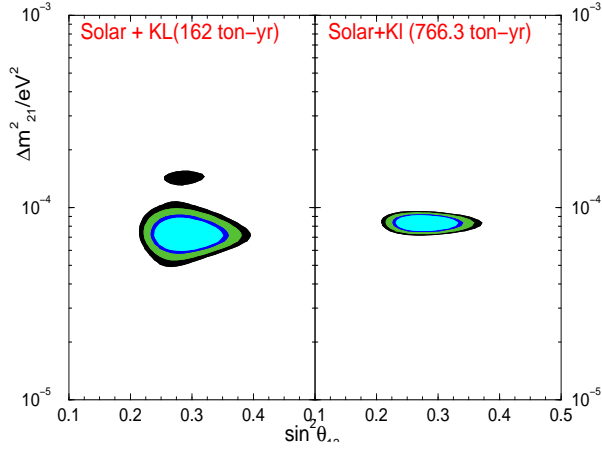}   
\caption{
Same as in figure 1
but from global
$\chi^2$-analysis of solar and \kl data.
}
\label{sol+kl}
\end{figure}
In  left(right) panel of \ref{sol+kl} we present the allowed 
regions from 
analysis of solar and  162 (766.3) ton year (Ty) \kl data.
In the left panel there are two allowed zones -- the low-LMA zone 
which is more favoured and the high-LMA zone admitted only at 3$\sigma$.
In the right panel
there is no allowed high-LMA region  
at 3$\sigma$ since 
the latest \kl data  disfavour higher $\dm$ values.  
The  allowed range of $\dm$  in low-LMA region 
also 
narrows  down considerably.
The best-fit points from combined analysis of solar 
with 162 Ty and 766.3 Ty \kl data are found 
respectively to be 
\\
$\bullet$ 
$\dm$ = 7.2 $\times 10^{-5}$eV$^2$,
$\sss$ = 0.29,
\\
$\bullet$
$\dm$ = 8.3 $\times 10^{-5}$eV$^2$,
$\sss$ = 0.27. \\
The $\theta> \pi/4$ ({Dark-Side})  solutions 
allowed by \kl spectrum data get disallowed by
the solar data as the matter effect in sun break the 
$\theta \rightarrow \pi/2 -\theta$ symmetry (cf. fig. \ref{prob}). 

\begin{figure}
\vskip -0.2cm
\includegraphics[width=9cm,height=5.8cm,clip=]{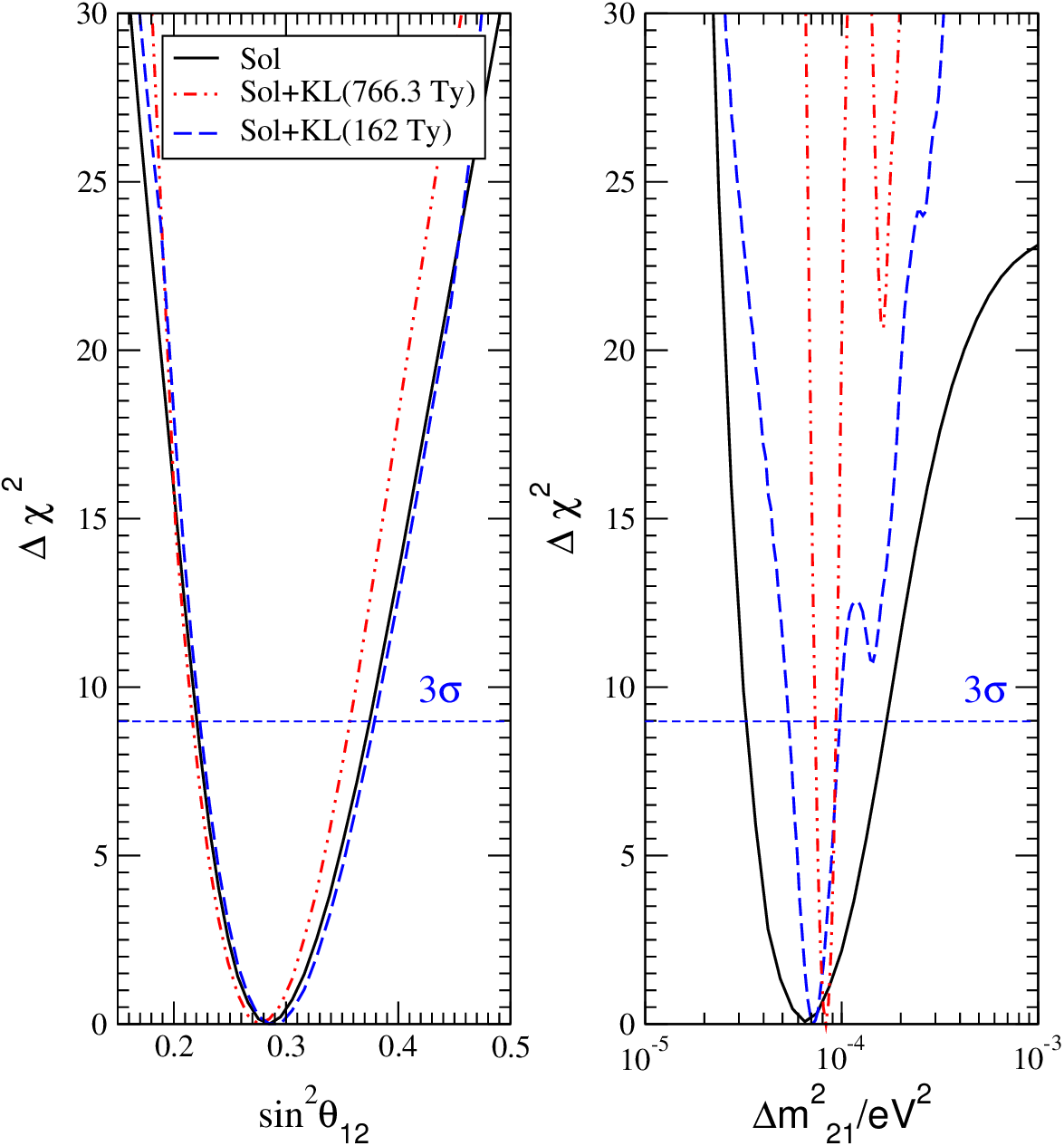}
\caption{\label{delvs12}
$\Delta \chi^2 $ vs \dm($\sss$)
marginalised over the other parameters. 
}
\label{delchi}
\end{figure}

In fig \ref{delchi} we plot $\Delta \chi^2 =\chi^2 -\chi^2_{min}$  
vs $\dm$ (right panel) and vs. $\sss$ (left panel).    
This figure shows that  with 
increased
statistics
\kl data constrains  
{$\Delta m^2_{21}$} more and more sharply.
However 
{$\sss$} is not constrained as  much. 
Maximal mixing is found to be more  disfavoured ($> 5\sigma$) 
by the new \kl data because of increased precision.  

The figure \ref{prob} shows the probabilities  
vs energy 
for KL (at an average distance of 180 km)
and solar neutrinos for  \dm =$ 7\times10^{-5}$ eV$^2$
and $\sin^2\theta_{12}$ =0.3.  
For these parameters 
the high energy  $^8B$  solar neutrinos 
undergo adiabatic MSW 
transition  with $P_{ee} \approx \sin^2\theta$. 
For low energy solar neutrinos  $P_{ee}  \approx 1 - 0.5 \sin^22\theta$.
Thus with decrease in energy the survival probability increases for 
$\theta < \pi/4$ causing  an upturn in the observed energy spectrum.  
Whereas for the solar probabilities the \dm dependence is completely averaged
out in KL
the probability exhibits a $L/E$ dependence which gives
it an unprecedented sensitivity  to  \dm.

The  
imprint of MSW effect \cite{msw} in Sun in the 
solar and \kl data can be quantified by parametrising  
the matter induced potential in the sun, 
{$V_{MSW} = \sqrt{2} G_F n_e$} , as
$\alpha_{MSW}.{V}$ \cite{lisimsw}.
$\alpha_{MSW}=0$ corresponds to no MSW effect
while $\alpha_{MSW}=1$ corresponds to full MSW effect.
Analysis performed in \cite{lisimsw}  with solar 
and 162 Ty \kl data shows that $\alpha_{MSW}=0$
case is rejected at 5.6$\sigma$ w.r.t the minimum which comes at 
$\alpha_{MSW}$ =1 i.e for standard MSW effect.     
However the allowed range of $\alpha_{MSW}$ 
was found to be large. 
Increased statistics from KL can put stronger constraints
on $\alpha_{MSW}$ and can provide
more precise test of {MSW} {and} "new" physics beyond MSW.
\\ \\
\underline{\bf Impact of each solar experiment}:
\begin{figure}
\includegraphics[width=8cm,height=6cm,clip=]{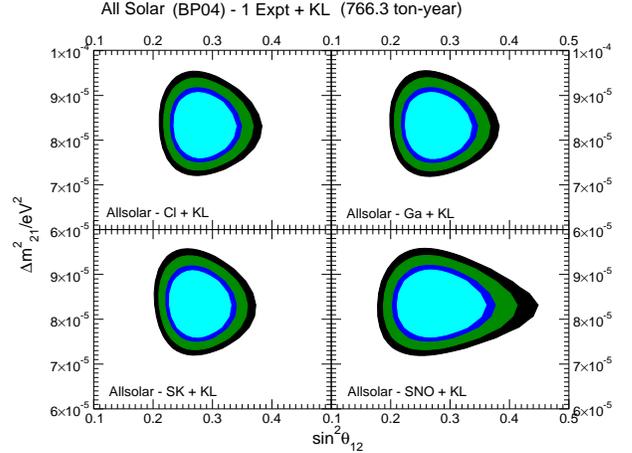}
\caption{\label{all-1}
The allowed regions in $\dm$-$\sin^2\theta_{12}$ plane,
obtained by removing the data from
one solar neutrino experiment from the global fit.
These figures are drawn on a linear scale for $\dm$.
}
\end{figure}
In Fig. \ref{all-1} we present the allowed
regions obtained by taking out
the data from one solar neutrino experiment
from the global data set.
Taking out the SNO data allows 
smaller values of $f_B$,
and hence larger values of
of $\sin^2\theta_{12}$  ($P_{ee} \sim f_B \sin^2\theta_{12}$).
With 162 Ty \kl data excluding SNO  
allowed maximal mixing, Dark side
and higher $\Delta m^2$ solutions.
With the spectral distortion observed in the 766.3 Ty \kl data these 
regions are ruled out with increased confidence even without SNO.
\\ \\
\underline{\bf On the precision of $\nu_{\odot}$
oscillation parameters}:
\begin{figure}[htb]
\vskip -0.5cm
\includegraphics[width=6cm,height=6cm,clip=]{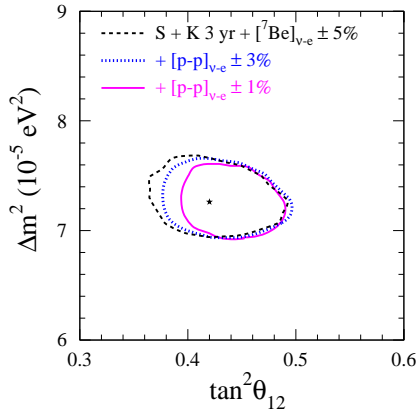}
\vskip -0.5cm
\caption{Allowed area  including  the rates from 
a generic pp neutrino experiment
with different errors. Figure taken from \cite{Bahcall:2003ce}}.
\label{ppfig}
\end{figure}
\begin{table*}[htb]
\caption{
3$\sigma$ allowed ranges and \% spread  of $\Delta m^2_{21}$ and
$\sin^2\theta_{12}$ obtained from 2 parameter plots. 
}
\label{spread}  
\newcommand{\m}{\hphantom{$-$}}
\newcommand{\cc}[1]{\multicolumn{1}{c}{#1}}
\renewcommand{\tabcolsep}{2pc} 
\renewcommand{\arraystretch}{1.2} 
\begin{tabular}{@{}lllll} 
\hline
{Data set} & (3$\sigma$)Range of & (3$\sigma$)spread in  &
(3$\sigma$) Range of
&(3$\sigma$) spread in \cr
{used} & $\Delta m^2_{21}$ eV$^2$
& {$\Delta m^2_{21}$} & $\sin^2\theta_{12}$
& {$\sin^2\theta_{12}$} \cr
\hline
{only sol} & 3.0 - 17.0
&{70\%} & $0.21-0.39$ &30\%\cr
{sol+162 Ty KL}& 4.9 - 10.7
& 37\%
& $ 0.21-0.39$ & 30\%  \cr
{sol+ 766.3 Ty KL}& 7.2 - 9.5
& 14\% &
$0.21-0.37$ & 27\% \cr
\hline
\end{tabular} \\[2pt]
\end{table*}
In  Table \ref{spread} we present the  3$\sigma$
allowed ranges of $\Delta m^2_{21}$
and $\sin^2\theta_{12}$, obtained
using different data sets.
We also show the uncertainty in the value of the
parameters through a quantity ``spread'' which we define in general as
\be
{\rm spread} 
= \frac{ \Delta m^2(\sin^2\theta)_{max} - \Delta m^2(\sin^2\theta)_{min}}
{\Delta m^2(\sin^2\theta)_{max} + \Delta m^2(\sin^2\theta)_{min}}
\ee
Table 1 illustrates the  
remarkable sensitivity of  \kl in reducing the uncertainty in
$\Delta m_{21}^2$.
But {$\theta_{12}$} is not constrained much better than the current
set of solar experiments. 
The reason for this 
is  the average energy and distance in \kl corresponds to 
a Survival Probability MAXimum
(SPMAX) i.e $\sin^2(\dm L/4E) \approx  0$. 
This means that the coefficient
of the $\sin^22\theta_{12}$ term
in $P_{\bar{e}{e}}^{KL}$ is relatively small,
weakening the  sensitivity of  \kl  to 
$\theta_{12}$.
The precision in $\theta_{12}$ can be improved
by reducing the baseline length such that one gets
a minimum of the
$\bar{\nu}_e$ survival probability (SPMIN) 
where
$\sin^2(\dm L/4E)= 1$. 
This corresponds to a distance 
$L = 1.24  (\mathrm E/MeV)
(\mathrm eV^2/\Delta m^2_{12})$ m.
For the low-LMA solution region
and the average energy of neutrinos in
\kl experiment,
this corresponds to a
distance of approximately (50 - 70) km \cite{th12}.
For an experiment with a 70 km baseline and 24.3 GW
reactor power
$\sss$ can be determined with 
$\sim 10\%$ error at 99\% C.L. with a 3 kTy 
statistics \cite{th12}.

Improved sensitivity in the measurement of $\theta_{12}$ 
is possible from LowNu experiments \cite{lownu}.
The {pp} flux is known with {1\%}
accuracy from Standard Solar Models as compared to 
the uncertainties of $^{8}{B} \sim$ {20\%} and $^{7}{Be} \sim$ {10\%}. 
At the low energies relevant for pp neutrinos 
$P_{ee}
\approx 1 - \frac{1}{2}\sin^22\theta$
resulting in a heightened sensitivity to $\theta_{12}$.
In \ref{ppfig} we show the allowed region  \cite{Bahcall:2003ce} in 
$\Delta m^2 -\tan^2 \theta_{12}$ plane including fictitious results 
from a generic pp scattering experiment. 
The figure shows that 
precision in $\theta_{12}$ increases with reduced error in pp 
flux measurment. 

The recently proposed Gd loaded SK detector  
can measure $\dm$ with
$\sim 1\%$ and $\sss$ with $\sim 15\%$ uncertainty at 99\% C.L.
after 5 years of data taking
\cite{skgd}.

\begin{figure}[htb]
\includegraphics[width=18pc]{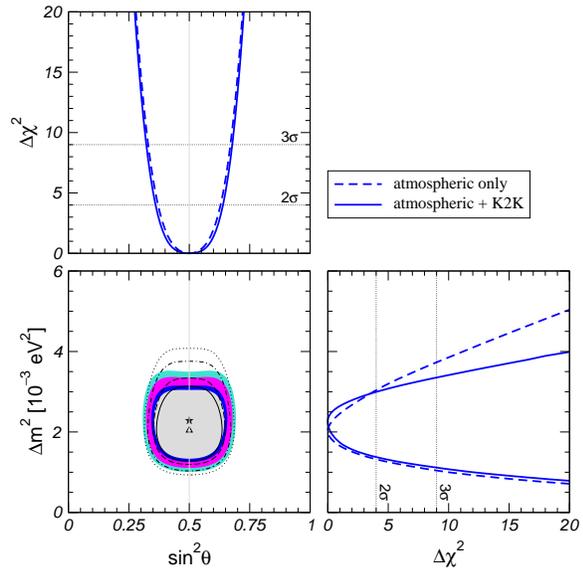}
\caption{The allowed regions from SK (line contours) and SK+K2K 
(shaded contours)
analysis from a
two flavour analysis.
Also shown is the  $\Delta \chi^2$ vs
$\Delta m^2_{atm}$ and $\sin^2\theta_{atm}$  marginalised 
w.r.t the undisplayed parameter\cite{maltoni}}.
\label{sk+k2k}
\end{figure}
\section{Atmospheric Neutrino Oscillation Parameters}
In this section we present the constraints  on atmospheric neutrino 
oscillation 
parameters $\Delta m^2_{32} \equiv \Delta m^2_{atm}$ and 
$\sin^2\theta_{23} \equiv \sin^2\theta_{atm}$ from
two flavour oscillation analysis 
 SK and SK+K2K 
data. 
\\ \\
\underline{\bf Allowed parameters from SK+K2K analysis}:
In 2003 the SK collaboration reanalyzed their data and reported a 
downward shift of the allowed $\Delta m^2_{atm}$ from a two-generation 
analysis
due to incorporation of 
three-dimensional Honda atmospheric fluxes and revised cross-section and
efficiencies. 
The best-fit value of $\Delta m^2_{atm}$ from 
SK zenith angle data reported in this conference is 
 2.1$\times 10^{-3}$
eV$^2$ \cite{kearns04}.
Analysis of SuperKamiokande data
by two groups confirm this downward shift of $\Delta m^2_{atm}$
\cite{Gonzalez-Garcia:2004it,maltoni}. 

The K2K  experiment which is sensitive to 
atmospheric neutrino oscillation parameters  
gives the best-fit as 
$\Delta m^2_{atm}$ = 2.8 $\times 10^{-3}$ eV$^2$ 
$\sin^22\theta_{atm}=1.0$.

In fig. \ref{sk+k2k} we show the allowed regions in 
$\Delta m^2_{atm}$ - $\sin^2\theta_{atm}$  plane from 
analysis of SK data as well as from combined analysis of 
SK+K2K data  
\cite{maltoni}.
The best-fit from the combined analysis corresponds to \\
$\bullet$
$\Delta m^2_{atm}$ = 2.3 $\times 10^{-3}$ eV$^2$, $\sin^2\theta_{atm}
= 0.5$. \\
Higher $\Delta m^2$ values are seen to be constrained
by K2K 
but $\theta_{atm}$ is not constrained any better. 
For a two generation analysis the relevant probability for 
both SK and K2K are 
vacuum oscillation probability and the allowed regions are symmetric about 
$\theta_{23}$ and $\pi/2 -\theta_{23}$. 

Recently SuperKamiokande has 
reported the observation of the first oscillation minima
in the $L/E$ distribution
\cite{ishi} .
The observation of this first dip disfavours the neutrino decay
and de-coherence solutions. 
The best-fit value of $\Delta m^2_{atm}$ from this analysis is 
2.4 $\times 10^{-3}$ eV$^2$ with 
$\sin^22\theta_{atm}$ = 1.0. 
The allowed ranges of parameters from the two different analyses are: 
\\
{$\bullet$}{90\% C.L. range}  (SK L/E) \\
{$\Delta m^2_{atm}$= {1.9} - 3.0 $\times 10^{-3}$ eV$^2$,
$\sin^22\theta_{atm} >$ 0.9 \\
$\bullet$
90\% C.L. range   (SK Zenith) \\
{$\Delta m^2_{atm}$= 1.3 - 3.0 $\times 10^{-3}$ eV$^2$,
$\sin^22\theta_{atm} >$ 0.9
\\ \\
\underline{\bf 
Precision of atmospheric neutrino oscillat-}
\underline{\bf ion parameters}:
The spread 
in $\Delta m^2_{atm}$ is 39\%    from SK zenith data 
whereas it is  22\%  from SK L/E data. 
Thus improved precision in {$\Delta m^2_{atm}$} is obtained 
with L/E data. Range of {$\sin^2\theta_{atm}$} however 
remains  unchanged: 
{$\delta(\sin^22\theta_{23}) \sim 5\%$. 
However  
$\delta(\sin^2\theta_{23}) \sim 32\%$ because   
{$\sin^2\theta_{23}$} precision is worse than
{$\sin^22\theta_{23}$} precision near maximal mixing. 
Increased statistics in
$L/E$ data can improve  
$\Delta m^2_{32}$ uncertainty to 
$\sim$ 10\%  at 90\% C.L. in  20 years of operation of SK \cite{kajitanoon}. 
This is sensitive to the {{true $\Delta m^2_{32}$}} chosen 
and the above precision is achieved  
for {$\Delta m^2_{32}$=2.5 $\times 10^{-3}$ eV$^2$}.
But precision in $\sin^2\theta_{23}$ does not improve
with increased statistics \cite{kajitanoon}. 
{Large Magnetized Iron} calorimeters 
for atmospheric neutrinos (e.g MONOLITH,INO)
\cite{monolith,ino} can improve the precision in both atmospheric 
parameters.  

\section{Three Flavour Oscillation}
The oscillation parameters for a three flavour analysis are 
two mass  squared   differences
$\Delta m^2_{21} = \Delta m^2_\odot$,~~~~
$\Delta m^2_{31} = \Delta m^2_{CHOOZ}$
$\simeq \Delta m^2_{atm} =
\Delta m^2_{32}$  
and three mixing angles which are usually denoted as  
$\theta_{12}$,$\theta_{23}$ and $\theta_{13}$  with the mixing matrix  U  
parameterized   
in the standard MNS form. 
In the limit 
$\Delta m^2_{21} << \Delta m^2_{32}$ as is indicated by the two 
generation analysis of atmospheric 
and solar neutrino data.
In this limit the   
atmospheric  probabilites depend on $\Delta m^2_{32}$,
$\theta_{13}$, $\theta_{23}$ and 
solar neutrino probabilites depend on
$\Delta m^2_{21}$, $\theta_{12}$, 
$\theta_{13}$. The CP violation phases  
can be neglected in this limit. 
Clearly for 
$\theta_{13} =0$ solar and atmospheric
neutrinos decouple. 
The probability for the 
CHOOZ reactor experiment depends on $\Delta m^2_{31}$ and 
$\theta_{13}$. 
\\ \\
\begin{figure}
\includegraphics[width=6cm,height=6cm,clip=]{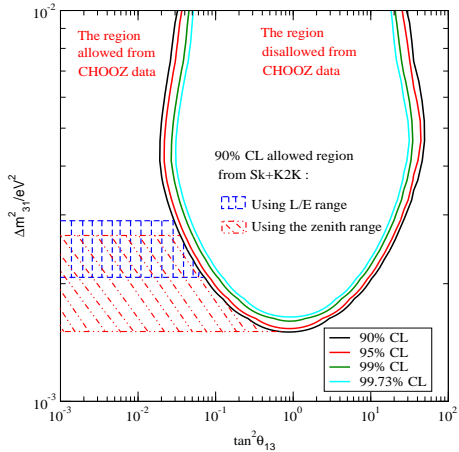}
\caption{Allowed area in $\Delta m^2_{31} - \tan^2\theta_{13}$ 
plane from CHOOZ data. 
The shaded regions correspond to the allowed range of 
$\Delta m^2_{31}$ using SK zenith angle (L/E) and K2K 
data \cite{Fogli:2003am,lisilbye}. }
\label{chooz} 
\end{figure}
\underline{\bf Bound on $\sin^2\theta_{13}$}:
CHOOZ/PaloVarde experiment has put stringent bounds on
$\theta_{13}$ from  non-observation of $\bar{\nu_e}$ 
disappearance \cite{choozref}. 
In fig \ref{chooz} we show the allowed area from CHOOZ experiment 
in $\Delta m^2_{31} - \tan^2\theta_{13}$ plane. 
The  area shaded by vertical blue (slanting red) lines 
correspond to the allowed range of $\Delta m^2_{31}$ 
from two generation analysis of SK L/E (zenith) + K2K
atmospheric neutrino data \cite{Fogli:2003am,lisilbye}. 
The figure shows that 
the {$\theta_{13}$} bound from CHOOZ 
depends sensitively on {$\Delta m^2_{31}$}.
Stronger bounds are obtained for  higher {$\Delta m^2_{31}$}.  

\begin{figure}
\includegraphics[width=6cm, height=6cm, clip=]{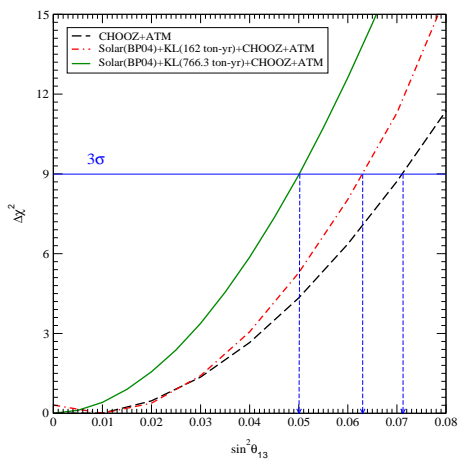}
\caption{\label{delchith13}
Bounds on the mixing angle $\theta_{13}$ from 
data combination. Also shown is the 3$\sigma$ $\Delta\chi^2$  limit for 
1 parameter. 
}
\end{figure}
In figure \ref{delchith13} we show the 
limits on $\theta_{13}$ obtained from   
$\Delta \chi^2$ vs 
$\theta_{13}$ plots. 
We 
marginalise over all other parameters.  
We let $\Delta m^2_{31}$ vary freely 
within the 3$\sigma$ range  obtained using the one parameter 
$\Delta \chi^2$ vs $\Delta m^2_{31}$ fit as given in the latest SK 
analysis \cite{kearns04}. 
The bounds that we get are \\
$\bullet$ $\sin^2{\theta_{13}} < 0.071 $ (CHOOZ+atm) \\
$\bullet$ $\sin^2{\theta_{13}} < 0.063 $ (sol+\kl(162)+CHOOZ+atm){\footnote{
Using the rage of $\Delta m^2_{31}$  from the SK+K2K analysis of
\cite{Fogli:2003am} bound comes as
{$\sin^2\theta_{13} < 0.077$}
\cite{snosaltus}}\\ 
$\bullet$ $\sin^2{\theta_{13}} < 0.05$ (sol+\kl(766.3)+CHOOZ+atm) \\
A non-zero $\theta_{13}$ prefers higher $\dm$  values which is disfavoured 
by the solar+\kl data. Because of this the combined $\chi^2$ increases 
making the bound on $\theta_{13}$ stronger than compared to 
only CHOOZ+atmospheric. Since the recent \kl data disfavours 
high $\dm$ values to a greater extent the bound on $\theta_{13}$ further 
improves by including it. 

In \ref{cont3g} we show the effect of  non-zero $\sin^2\theta_{13}$ on 
the allowed area in solar neutrino parameter space. 
The presence of a small non-zero $\theta_{13}$ can
improve the fit in the regions of
the parameter space with higher values of $\dm$  \cite{snosaltus},
i.e., in the high-LMA zone. But due to increased discord between 
the new \kl data  and high $\dm$ values 
the
high-LMA region now gets excluded at more than
 3$\sigma$ even in the presence of a
third generation in the mixing,
indicating the robustness of the low-LMA solution.
The allowed atmospheric regions in $\Delta m^2_{31} - 
\sin^2 2\theta_{23}$ plane also remains stable in presence of 
non-zero $\theta_{13}$ \cite{maltoni}. 
Including the $\Delta m^2_{21}$ terms in the analysis one can give 
bound on the 
hierarchy  parameter $\alpha = \Delta m^2_{21}/\Delta m^2_{31}$
associated with subleading oscillations  for atmospheric neutrinos 
as well as for long baseline studies. 
The best-fit for this comes as  
$\alpha$ = 0.03
\cite{maltoni}.
\begin{figure}[htb]
\vskip -0.2cm
\includegraphics[width=6cm,height=6cm,clip=]{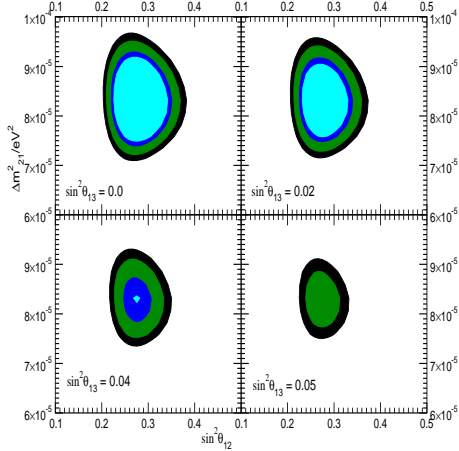}
\caption{\label{solkl3osc}
Same as in figure 1 but from
a three flavour analysis of the
global solar and reactor neutrino data and plotted on a linear scale for $\dm$.
}
\label{cont3g}
\end{figure}

\section{Conclusions}
A series of seminal experiments performed over a period of
nearly four decades and more and more refined analysis of 
the global experimental data has not only 
confirmed the existence of neutrino mass and mixing,
but also narrowed down the allowed parameter ranges 
considerably heralding the  precision era in neutrino physics. 

The prime future goals for neutrino oscillation studies are- 
further improvement of precision  of the parameters,
to obtain some measure of smallness of {{$\theta_{13}$}}
, to determine the {{sign of $\Delta m^2_{32}$}} and to
search for
{CP violation} in the lepton sector. 
\\ \\
{\bf{Acknowledgements}}:S.G. would like to thank the organizers  of 
Neutrino 2004 for the invitation to give this talk and 
T. Kajita, E. Lisi and M. Maltoni for many helpful discussions during the 
preparation of the talk.


\end{document}